\documentclass[aip, 
 amsmath,amssymb,
 reprint,%
floatfix,
]{revtex4-1}

\usepackage{graphicx}
\usepackage{dcolumn}
\usepackage{bm}

\usepackage[utf8]{inputenc}
\usepackage[T1]{fontenc}
\usepackage{mathptmx}
\usepackage{etoolbox}
\usepackage{float}
\usepackage{xcolor}
\usepackage{comment}

\makeatletter
\def\@email#1#2{%
 \endgroup
 \patchcmd{\titleblock@produce}
  {\frontmatter@RRAPformat}
  {\frontmatter@RRAPformat{\produce@RRAP{*#1\href{mailto:#2}{#2}}}\frontmatter@RRAPformat}
  {}{}
}%

\makeatother
\begin{document}

\preprint{AIP/123-QED}

\title{Femtosecond dynamics on the nanoscale of intense laser-induced grating plasma}

\author{Ankit Dulat} 
\author{Sagar Dam}
\author{Sk Rakeeb}
\author{Amit D. Lad}
\author{Yash M. Ved}
\author{G. Ravindra Kumar$^{*}$}
\email{grk@tifr.res.in}

\affiliation{Tata Institute of Fundamental Research, Dr. Homi Bhabha Road, Colaba, Mumbai-400005, India}

\begin{abstract}

The complex interaction dynamics of intense femtosecond (fs) pulses and their picosecond (ps)-long leading edge with nanostructured solids occur at both the nanometer and the femtosecond scales, making them extremely difficult to measure directly. Here, we present pump-probe-based measurements that capture the ultrafast evolution of relativistically intense laser-driven grating plasma on fs time and nanometer spatial scales. We measure the transient reflectivity and spectrum of the scattered or diffracted UV-probe pulses from the grating structures with 100s of fs resolution. Our measurements capture the initial onset of the solid-to-plasma transition and the subsequent grating plasma expansion, a few ps before the peak of the intense fs pulse. We measure the instantaneous position of the electron critical surface, its velocity, and its acceleration, which are very crucial for fundamental understanding and applications in ion/electron acceleration and high harmonic generation, while also providing valuable benchmarks for simulations. Particle-in-cell (PIC) simulations corroborate the observations well offering further insight into this process.
\end{abstract}

\maketitle

\textbf{\textit{Introduction}} -- Ultrahigh-power femtosecond laser pulses interacting with nanostructured solids create extreme energy-density states ($>10^{9}$ J/c$m^3$) \cite{Hollinger2020,Bargsten2017,Kaymak2016,Purvis2013} achievable in the laboratory. These interactions pave the way for tabletop-scale fusion \cite{Curtis2018,Curtis2021}, miniaturized particle accelerators \cite{Fedeli2016,Lad2022,Moreau2020,Chatterjee2012,Jiang2016,Ceccotti2013,Zigler2013}, and powerful extreme-ultraviolet (XUV) light sources \cite{Hollinger2017,Mondal2011,Bagchi2011,Rajeev2003,Rajeev2002}. Furthermore, recent studies with nanostructured solids have demonstrated spatio-temporal control of relativistic electrons \cite{Dulat2024} and XUV light \cite{Cerchez2013,Yeung2013} generated by these extreme interactions, offering significant potential for the development and optimization of these applications. 


However, the development of next-generation laser-based applications is strongly hindered by the lack of experimental ability to probe the electron density dynamics on fs and nanometer scales. This includes understanding the interaction of the ps-long pedestal and the steep rising edge in intensity before the peak of the fs pulse (leading edge interaction) in petawatt and multi-petawatt lasers. The intensity of the laser field at the rising edge increases by many orders of magnitude in just a few ps, resulting in the ionization of the target, forming a so-called pre-plasma \cite{Adumi2004}. The interaction of the peak of the fs pulse critically depends on the pre-plasma state and, in particular, on the scale length of its density gradient.

The density gradient critically determines the laser absorption mechanism \cite{Gibbon2005}, energy transfer to electrons \cite{Santala2000,Macphee2010,Ovchinnikov2013}, and the transition between different mechanisms of ion acceleration \cite{Culfa2017,Machi2013}. High-harmonic generation efficiency is extremely sensitive to even small variations in the density gradient \cite{Tarasevitch2007,Thaury2010}, on the scale of only a few tens of nanometers, i.e., a small fraction of the laser wavelength $\lambda_L$. The effects of pre-plasma expansion are even more critical for the nanostructured target, as its original structure can be completely altered and strongly affects the interaction mechanism. Therefore, it is crucial to accurately measure the electron density dynamics during the rising edge and the peak of the fs pulse, extremely challenging to do on nanometer and femtosecond scales.

Simulating pre-plasma dynamics requires sophisticated numerical modeling, which often relies on simplified assumptions due to the inherent complexity of the problem. Additionally, plasma formation is critically dependent on the intricate temporal variation of laser intensity at the leading edge over a wide dynamic range. As a result, simulations cannot be fully predictive and necessitates complementary experimental investigations.

Experimentally, only a few measurements \cite{Blanc1996,Dulat:22,Kluge2018,Randolph2022,Malvache2013} have been capable of capturing the fs evolution of pre-plasma with nanometer spatial resolution for 10s of ps duration. The most promising is pump-probe-based small-angle x-ray scattering (SAXS) \cite{Kluge2018,Randolph2022,Malvache2013} using x-ray free-electron laser (XEFL) probe pulses. However, such a technique requires large XEFL facilities and is not accessible in table-top experiments. Furthermore, the scattered x-rays probe the dynamics both at the critical surface and the overdense plasma region beyond it. This complicates the interpretation of the data, as the crucial laser-plasma interaction occurs primarily within the skin depth of a few nanometers at the critical surface. Therefore, a method capable of selectively sampling the dynamics near the critical surface is more suitable. Frequency domain interferometry (FDI)-based measurements \cite{Blanc1996} are challenging due to stringent requirements, including extreme vibration stabilization. Furthermore, in FDI, various factors contribute to the measured phase, making data interpretation very difficult and limit its applicability.

\begin{figure}[ht]
\centering
\includegraphics[width=1\linewidth]{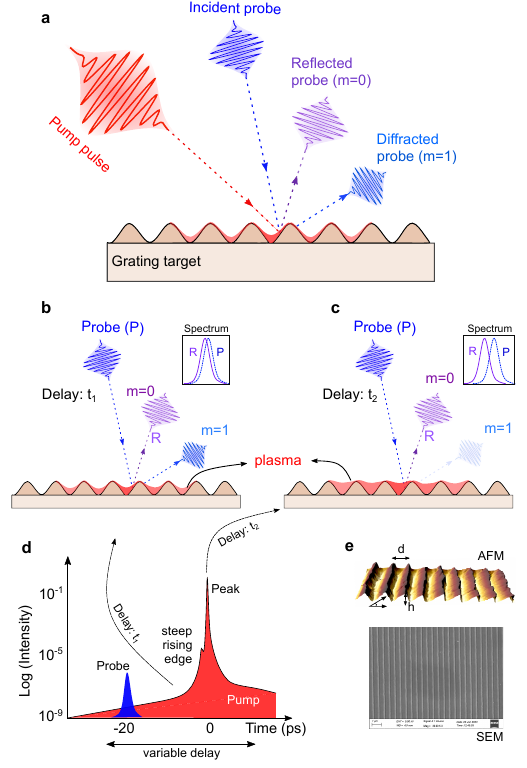}
\caption{(a) Schematic of the pump-probe setup. (b-c) Schematic showing the effect of pre-plasma expansion on the diffraction efficiency of the diffracted probe in first-order mode (m = 1) and the spectrum of the reflected probe pulse (m = 0) for two different pump-probe delays. (d) Picosecond intensity profile of the fs pump pulse. (e) AFM and SEM images of the gold-coated triangular blazed grating used in the experiment (Jobin Yvon, 1800 lines/mm groove density, 555-nm period, 158-nm groove depth, $17.5^\circ$ blaze angle).}\label{fig: Fig1}
\end{figure}
Here, we present the first measurement of the fs dynamics of a grating pre-plasma driven by a relativistic intense laser pulse using a much simpler and more direct method. Using the fs pump-probe technique, we measure the transient reflectivity and spectrum of the scattered or diffracted UV probe pulse from the transient plasma on the grating. By monitoring the time-dependent reflectivity and Doppler shifts in the fs probe beam, we are able to infer the ultrafast evolution of the electron density profile. We present the clear onset of the solid-to-plasma transition and the subsequent grating-plasma expansion by the rising edge and the peak of the intense fs pulse. Furthermore, we derive the instantaneous location of the electron critical surface, its velocity, and acceleration profiles. Our experimental results are consistent with the results of 2D electromagnetic particle-in-cell (PIC) simulations \cite{Arber:2015hc}. We believe our results are very important for a fundamental understanding of electron acceleration and HHG from grating targets at relativistic intensities, while also providing valuable benchmarks for simulations.

\textbf{\textit{Experiment}} --
The experimental setup and the working principle of the measurement are shown in Figure 1. A p-polarized pulse from the tabletop Ti:sapphire chirped pulse amplifaction laser (25 fs, 800 nm, 150 TW maximum) at the Tata Institute of Fundamental Research is used as a pump beam and is focused at a $40^\circ$ angle (Non-SPR angle) of incidence (AOI) onto the grating target, using an off-axis parabola. The measured focal spot diameter of 8 $\mu$m gives peak intensities of 5$\times10^{18}$ W/c$m^2$ at the pulse energy used in the experiment. Fig. 1e shows the atomic force microscopy (AFM) and scanning electron microscope (SEM) images of the gold-coated triangular blazed grating target used in the experiment. The pump pulse has a picosecond contrast (ratio of intensity at the peak to the leading edge) of $10^9$  at 25 ps before the peak of fs pulse (Fig. 1d). A small fraction of the pump pulse is frequency doubled by a 2 mm-thick $\beta$-barium borate (BBO) crystal to generate a probe pulse at 400 nm (shown in blue in Fig. 1). The probe pulse is s-polarized and has a pulse duration of 80 fs (measured using SD-FROG \cite{Aparajit2023}). A BG-39 filter is used to remove the residual 800 nm present in the probe beam. The time delay between the pump and probe is varied using a delay stage of 1 micron resolution. The probe pulse is incident at an AOI of $5^\circ$ and is focused at a spot of 60 $\mu$m, which is spatially overlapped with the pump focal spot. Precise measurement of "time-zero," i.e., temporal overlap of the peak of the fs pump and the probe pulse, is very crucial. Time-zero was measured by monitoring the reflectivity of the probe pulse from a test glass target as a function of pump-probe delay. At zero time delay, there is a sharp increase in the reflectivity of the glass plasma (see Fig. \ref{fig: Fig4} in Appendix A). The pump intensity was reduced to $10^{15}$ W/c$m^2$  (for time-zero measurement) to ensure no pre-plasma is generated by the leading edge of the pulse.

The incident probe pulse is diffracted by the grating target, and the diffraction angle is given by $\theta_d$ = $\sin^{-1}(\sin\theta_i \pm m\lambda/d)$, where $\theta_i$ is the AOI of the laser, m is the diffraction order, and d is the period of the grating structure. As shown in Fig. 1a, m = 0 corresponds to specular reflection, while m = 1 corresponds to first-order diffraction at $\theta_d$ = $60^\circ$. The diffraction efficiency of the grating structure depends primarily on the material properties, the exact topography of the groove structure, and, more specifically, the depth of the groove. We monitored the transient reflectivity of the probe pulse in both the reflection and first-order diffraction modes as a function of pump-probe delay. For our grating target, the diffraction efficiency in the first-order mode is extremely sensitive to even a small change in the groove depth, on the order of a few nanometers. Thus, monitoring even the smallest changes in the diffraction efficiency in first-order mode allows us to infer the dynamics at the nanoscale. Simultaneously, Doppler shifts in the reflected probe (m = 0) reveal the instantaneous velocity of the hydro-expansion of the pre-plasma. The schematics in Figs. 1b and 1c show the evolution of both first-order diffraction efficiency and Doppler shifts in the reflected probe (m = 0) as the pre-plasma, induced by the rising edge of the pump pulse evolves on femtosecond and nanometer scales. The 400 nm probe pulse has a narrow spectrum ($\sim$ 1 nm) with a nearly Gaussian profile (unlike pump pulses with bandwidths as large as 60 nm, and a complex spectral profile), making it easy to characterize even small Doppler shifts. Secondly, it vastly improves the signal-to-noise ratio because all pump noise can be removed with a narrow spectral filter around 400 nm.\\

\textbf{\textit{Results}} -- Figure 2 depicts the experimental results obtained with a pump pulse having a peak intensity of $5\times10^{18}$ W/c$m^2$. Fig. 2a shows the time-delayed reflectivity of the probe pulse from the transient plasma on the grating along the specular (m = 0) and first-order diffraction (m = 1) modes. The negative delays represent the plasma dynamics driven by the leading edge of the pump pulse (i.e., the probe pulse precedes the peak of the fs pump pulse). The grating target (without plasma) couples about 55$\%$ of incident light into zeroth-order mode, while 28$\%$ is in first-order mode, and the rest is in m = -1 mode. For time delays before -4 ps, we observe no change in the reflectivity of the probe for both zeroth and first-order modes, and the reflectivity is the same as if there were no plasma. However, around -3 ps, we observed a fall in the reflectivity of the diffracted mode and a slight increase in the reflectivity of the specular mode. This change in reflectivity indicates the onset of ionization of the grating target, as discussed in detail later. Reflectivity in diffracted modes continues to rapidly decrease and is reduced to merely 10$\%$ around the peak of the pump pulse, which later almost reaches zero (below the sensitivity of our measurement) after 500 fs of the peak. On the other hand, the reflectivity of the specular mode increases from 55$\%$ to 70$\%$ during the negative delays. In the positive delays, however, we observe a sharp drop (70$\%$ to 10$\%$) in the reflectivity that saturates around 8$\%$ at about 2 ps. 

Figure 2b shows the simultaneously measured spectrum of the reflected probe in the zeroth-order mode. We observe a slight blue shift that progressively increases from -4 ps to 0.6 ps. 
\begin{figure}[t]
\centering
\includegraphics[width=0.9\linewidth]{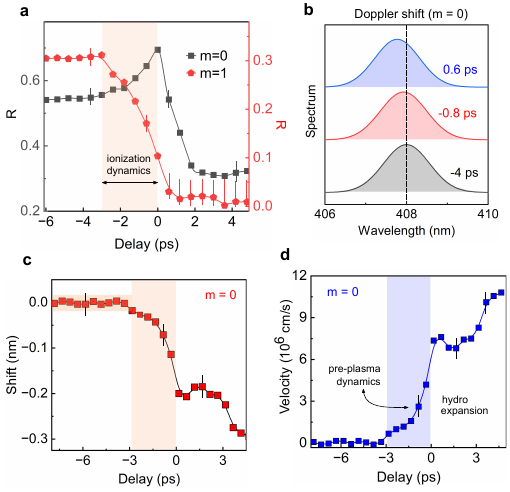}
\caption{(a) Time-delayed reflectivity of the probe pulse from the transient grating plasma along the specular (m = 0) and first-order diffraction (m = 1) modes. (b) Spectrum of the reflected probe pulse at three different time delays. (c) Time-delayed Doppler shifts induced in the central frequency of the probe pulse by the pre-plasma expansion. (d) The measured expansion velocity of the critical surface for 400 nm probe pulse.}\label{fig: Fig2}
\end{figure}
Figure 2c shows the time-delayed Doppler shifts induced in the central frequency of the probe pulse by the pre-plasma expansion. We observe a very small blue shift starting about 2.5 ps before the peak of the pump pulse. This shift rapidly increases as the pump pulse peak approaches and continues to rise for several ps after the peak. The observed blue shifts indicate that the plasma critical surface is expanding towards the vacuum. The expansion velocity can be calculated from the measured Doppler shifts using the equation v = $0.5c\times \Delta\lambda/(\lambda \cos\theta)$ \cite{Mondal2010,Jana2021}, where $\Delta \lambda$ is the measured wavelength shift, $\lambda$ is the central wavelength of the incident probe pulse, and $\theta$ is the angle of incidence. Figure 2d plots the calculated velocity for various time delays. We observe that the plasma surface accelerates during the rising edge of the pump pulse and continues to accelerate even after the peak of the femtosecond pulse. The average acceleration near the peak of the pulse reaches a value of $10^{18}$ cm/$s^2$.

\begin{figure}[t]
\centering
\includegraphics[width=1\linewidth]{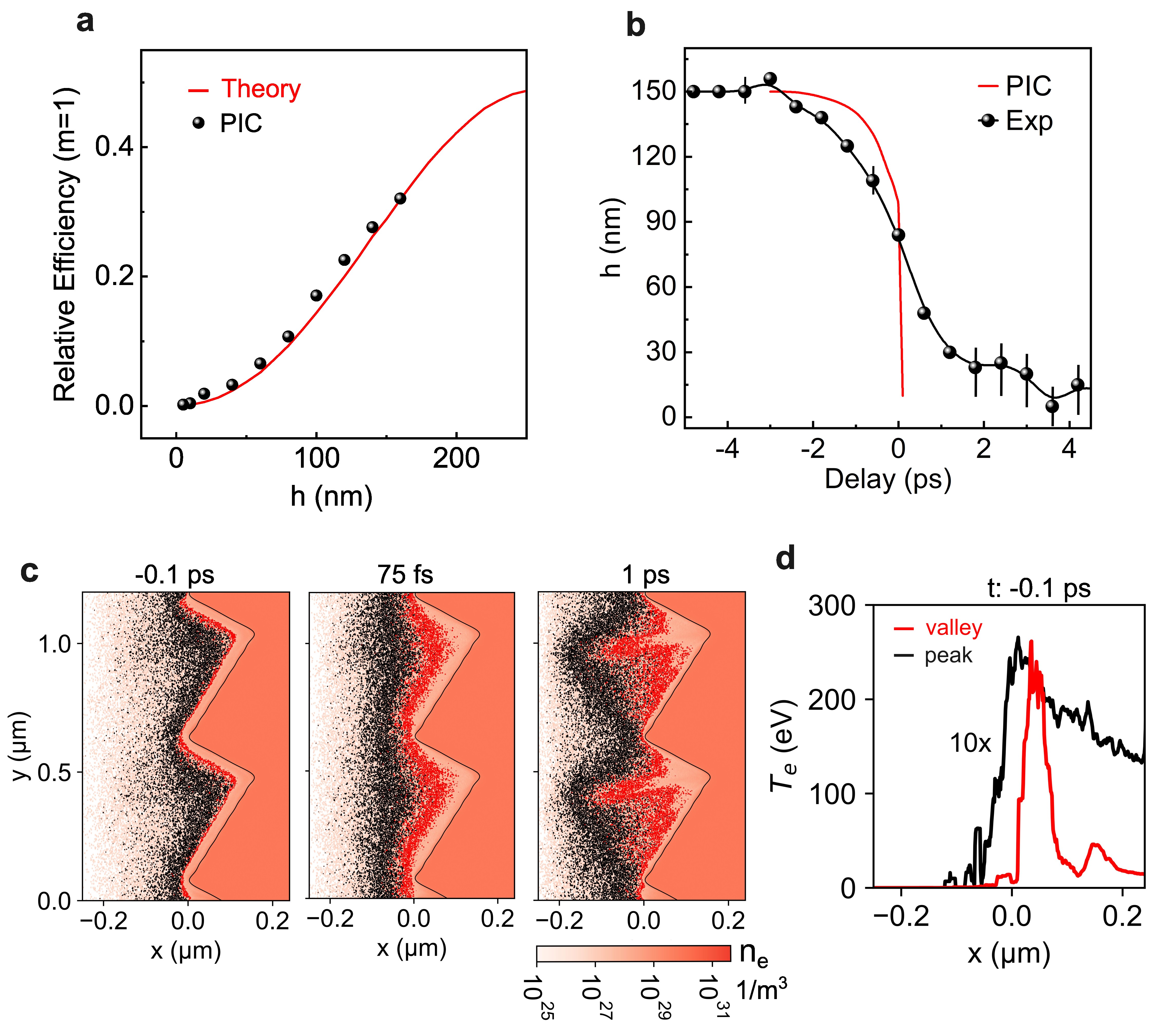}
\caption{(a) Relative diffraction efficiency (along m=1 mode) of the transient plasma grating for different groove depths (b) Sub-picosecond changes in the groove depth of the grating due to pre-plasma expansion. (c) Femtosecond evolution of the electron density profile of the grating structure obtained from PIC simulations. The black and red scatter points represent the critical surfaces of the pump and probe pulses, respectively. A thin black line outlines the boundary of the initial grating structure. (d) Temperature profile in the peak (black) and the valley (red) of the grating grooves, 100 fs before the peak of the pump pulse hit the grating target. The black curve is multiplied by a factor of 10.}\label{fig: Fig3}
\end{figure}
To quantitatively understand pre-plasma expansion within the grating grooves, the diffraction efficiency of the probe pulse was modeled using both theory\cite{Hutley} and PIC simulations. Figure 3a shows the calculated diffraction efficiency as a function of the change in grating groove depth (h) for an AOI of $10^{\circ}$. The theoretical calculations assume a very steep density gradient at the vacuum-grating interface and that the probe reflectivity does not change with groove depth. PIC simulations, on the other hand, incorporate the effect of a finite density gradient arising from the hydrodynamic expansion of the grating plasma and align well with the theory. Using the experimental data in Fig. 2a and the calculated diffraction efficiency (Fig. 3a), we determined the change in grating groove depth due to the hydrodynamic expansion of the grating plasma (shown in Fig. 3b). We observe a decrease in groove depth, starting approximately 2.5 ps before the peak of the fs pulse. Near the pulse's peak, the depth reduces to roughly 80 nm from its initial 155 nm—a decrease of approximately 50$\%$. This significant alteration in groove depth can profoundly impact the interaction mechanism and, consequently, particle acceleration.

To gain further insight into the interaction, we employed the EPOCH code \cite{Arber:2015hc} to simulate the hydrodynamic expansion of the grating plasma. Simulation parameters closely matched the experimental setup (details provided in the Appendix). Figure 3b shows the simulated evolution of groove depth (red curve) along with the experimental data. While qualitatively consistent, the simulation underestimates pre-plasma formation at negative time delays. This can be attributed to two simplifications: (1) the simulation only ran from -2.5 ps to 1 ps, neglecting any ionization occurring before -2.5 ps; and (2) collisional ionization calculations were performed every 50 steps (approximately 150 as). Additionally, deviations at positive time delay can be attributed to the fact that in the experiment, the probe focal spot is larger than the pump spot, and hence, a part of the probe was sampling the unaffected region.

In the simulations, we observe an interesting effect: the formation of a transient plasma grating structure ahead of the preinscribed grating due to plasma expansion approximately 1 ps after the pulse's peak. Figure 3c illustrates the evolving electron density profile of the expanding plasma at -100 fs, 75 fs, and 1 ps. The black and red scatter plots mark the critical surfaces of the pump and probe pulses, respectively. We observe plasma jets originating from each groove valley. These jets expand and coalesce, forming a transient plasma grating (with critical density modulation) in an intermediate position. This effect mirrors the optical Talbot effect, where, in spite of the light field, electron density modulation is repeated at a distance away from the grating plane. A similar phenomenon has previously \cite{Kluge2018} been observed with a square-shaped grating target. To better understand this effect, Fig. 3d shows a line cut of the background plasma temperature profile (ignoring fast electrons) along the tip and valley of a grating groove. Notably, the temperature near the valley is approximately 10 times higher than at the tip, revealing significantly inhomogeneous heating of the structure. This inhomogeneous temperature profile sets up plasma expansion with different velocities at the tip and valley, resulting in the generation of the transient plasma grating. Simulations indicate expansion velocities as high as $2\times10^7$ cm/sec, which is closely aligned with the experimental measurements. The inhomogeneous heating of the grating is due to the inhomogeneous localization of the laser field in the grating grooves, as shown in Fig. 5 of the Appendix. 

\textbf{\textit{Conclusion}} -- 
To conclude, we have performed experiments and simulations to measure the ionization dynamics and the hydrodynamical expansion of the pre-plasma driven by the steep-ps rising edge and the fs peak of a relativisitic laser pulse on a grating target. We present electron density dynamics with fs temporal and nanometer spatial resolution with a much simpler and direct pump-probe-based method. We quantitatively present the extent of pre-plasma expansion and its velocity profile. We emphasize that these measurements are extremely crucial for understanding HHG generation and particle acceleration from the grating target. Nevertheless, the pre-plasma measurements on the nanoscale can also serve as a sensitive indicator of the picosecond contrast of relativistic laser pulses.
	
\section*{ACKNOWLEDGMENTS} 
 GRK acknowledges partial support from J.C. Bose Fellowship grant (JBR/2020/000039) from the Science and Engineering Board (SERB), Government of India. Simulations were performed using EPOCH, which was developed as part of the UK EPSRC grants EP/G054950/1, EP/G056803/1, EP/G055165/1 and EP/ M022463/1.The authors acknowledge the TIFR HPC resources used for the simulations reported in this paper. We would like to acknowledge Ameya Parab for his help during the experiment.
	
\section*{DATA AVAILABILITY} The data that support the findings of this study are available within the article.

\appendix
\section{Time zero measurement}
\begin{figure}[h]
\centering
\includegraphics[width=0.6\linewidth]{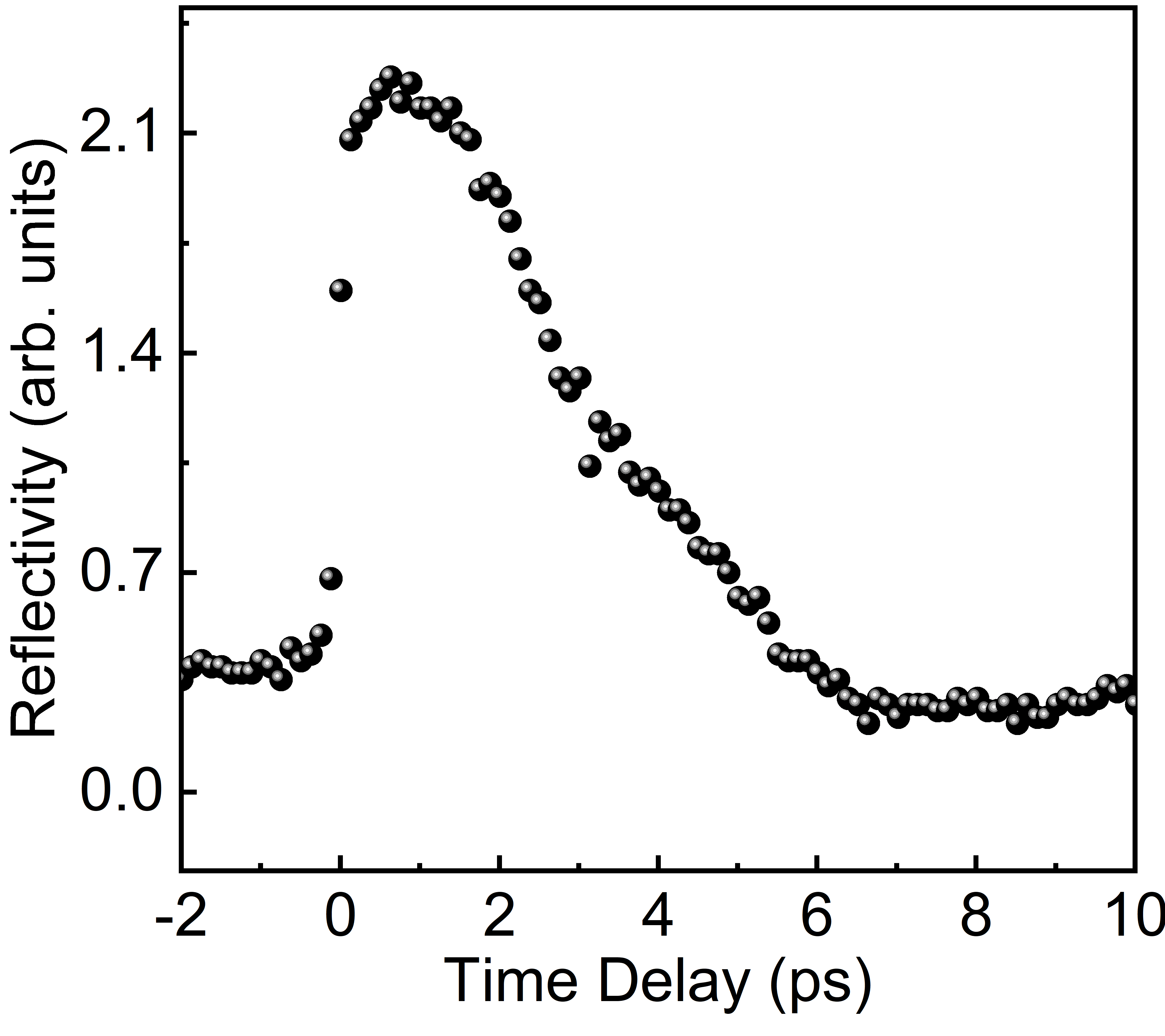}
\caption{Time-resolved reflectivity of the probe pulses from a fused silica target that is excited by
a pump pulse of intensity $\sim 10^{15}$ W/c$m^2$}\label{fig: Fig4}
\end{figure}

\section{PIC Simulation}
We performed 2D3v particle-in-cell (PIC) simulations in Cartesian geometry using the EPOCH code \cite{Arber:2015hc}. The simulation box size is 0.5$\mu m \times 1.2\mu m$ with a cell size of 1nm $\times$ 2nm and 64 macroparticles per cell. The target was modeled as a gold triangular blazed grating with peridodic boundary conditions along the y axis. The grating parameters, with a period of 555 nm, a groove depth of 158 nm, and a blaze angle of $17.5^\circ$, were used (similar to the experiment). The grating plasma was initialized with singly ionized gold ($Au^+$) ions and neutralizing electrons, all at an initial temperature of 25 meV. To model the ionization of the grating target, collisional ionization, multiphoton ionization, and barrier-suppression ionization modules were included. A p-polarized laser pulse with a central wavelength of 800 nm irradiates the vaccum-plasma interface at an AOI of 40°. The laser pulse is assumed to be Gaussian in both the longitudinal and transverse directions, with a FWHM pulse duration of 25 fs and a beam waist of 2 $\mu$m at the focus. The peak intensity of the focused laser is 5× $10^{18}$ W/c$m^2$, which is the same as the intensity used in the experiment. The temporal profile of the laser pulse was modeled with a picosecod pedestal and a steep rising edge preceding the peak of the fs pulse (very similar to the laser profile shown in Fig. 1d). Simulations were run for 3.5 ps of duration. To reduce computational cost and speed up the code, collisional calculations were performed every 50 steps ($\sim$150 attoseconds).

\section{Near-field Distribution}
\begin{figure}[h]
\centering
\includegraphics[width=1\linewidth]{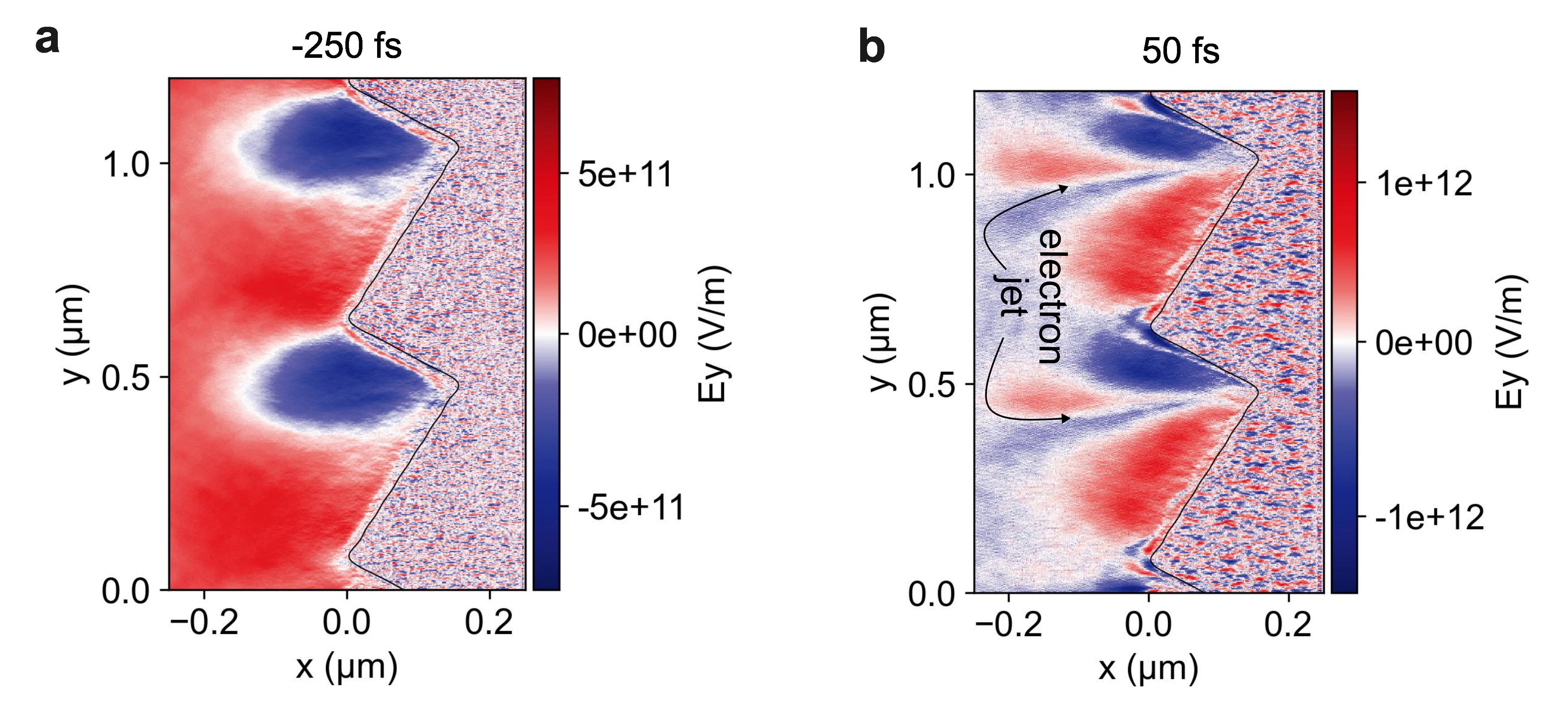}
\caption{(a) and (b) show the distribution of the y-component of the laser electric field ($E_y$) near the grating grooves at -250 fs and 50 fs time delays, respectively. Notably, in (b), we observe electron jets emnating from the valley of the groove.}\label{fig: Fig5}
\end{figure}
\bibliography{reference}

\end{document}